\documentstyle[prb,twocolumn,aps]{revtex}
\newcommand{\be}{\begin{eqnarray}}
\newcommand{\ee}{\end{eqnarray}}

\begin{document}

\title{Activated Dynamics
and the Ergodic-Nonergodic Transition}
\author{Gene F. Mazenko }
\address{The James Franck Institute and the Department of Physics\\
The University of Chicago\\
Chicago, Illinois 60637}
\date{\today}
\maketitle
%
%
\begin{abstract}

The hindered diffusion model is introduced. It is a continuum model giving
the dynamics of a conserved density.  Similar to the spin-facilitated models,
the kinetics are hindered by a fluctuating diffusion coefficient that decreases
as the local density approaches some geometrically constrained value
where the mobility goes to zero.  The model leads in a natural way to
activated dynamics at low temperatures and high densities.
In a well defined approximation the theory is shown to be 
compatible with mode-coupling
theory.  Indeed in the simplest form of the model we find an 
ergodic-nonergodic transition
into a phase that supports activated kinetics and nonergodic behavior.

\end{abstract}


We introduce a simple model for the structural glass transition
that naturally leads to activated dynamics at low temperature and high densities.
This model is  continuum generalization of the spin-facilitated (SF)
dynamics\cite{FA,chan,WBG1,jack}
that have been put forth as models for
glassy dynamics.  It is more physical than the SF models since it is formulated
in terms of a density field $\rho $ that is conserved.
As in the SF models, the equilibrium static behavior can be chosen
to be noninteracting.  In the continuum description this corresponds to
choosing Gaussian static
statistics. 
The model can also be used to make contact with mode-coupling theory\cite{MCT,BB}.
The key new ingredient in this approach is to choose a bare
diffusion coefficient\cite{coll} that is density dependent $D(\rho )$.  Technically
this complicates the
problem compared to the simplest field theoretical models that have been
proposed\cite{DM,DR}.
This is because we must
treat multiplicative noise.  Generally this
has been a difficult proposition.  However, there are certain constrained models,
typically with conservation laws, where the treatment\cite{KM91,Morozov} 
of multiplicative noise
is tractable.
The general class of models includes those discussed by Dean\cite{Dean},
Kawasaki and Miyazima\cite{kaw}, and Miyasaki and Reichmann \cite{RFT}. The important
difference is that they deal with a bare diffusion coefficient that is
linear in the density.
We study here the situation where there is a high density constraint on the
bare diffusion coefficient.

The model we study here is given in Langevin
equation language by
\be
\frac{\partial \rho}{\partial t}
=\vec{\nabla}\cdot\left(D(\rho)\vec{\nabla}
\frac{\delta {\cal H}_{\rho}}{\delta\rho}\right)
+\vec{\nabla}\cdot\left(g_{\rho}\vec{\eta}\right)
\nonumber
\ee
where $D(\rho )$ is the bare diffusion coefficient defined below,
$\vec{\eta}$ is Gaussian-white noise with variance
\be
\langle\eta_{\alpha} ({\bf x},t)\eta_{\gamma} ({\bf y},t^{\prime})\rangle
=2k_{B}TD_{0}\delta_{\alpha\gamma}
\delta({\bf x}-{\bf y})\delta (t-t^{\prime})
\label{eq:2}
~~~,
\ee
the multiplier 
$g_{\rho}=\sqrt{\frac{D(\rho)}{D_{0}}}$
and 
the effective Hamiltonian ${\cal H}_{\rho}$ can be taken to be quadratic
in $\rho$:
\be
{\cal H}_{\rho}=\int ~ d^{d}x_{1}d^{d}x_{2}\frac{1}{2}
\delta\rho ({\bf x}_{1})\chi^{-1} ({\bf x}_{1}-{\bf x}_{2})\delta\rho ({\bf x}_{2})
\nonumber
\ee
where $\delta\rho ({\bf x}_{1})=\rho ({\bf x}_{1})-\rho_{0}$.

The field theoretic version of this model corresponds to the
MSR\cite{MSR} action
given by
\be
A=\int  d^{d}x dt\left[ D(\rho )(\nabla \hat{\rho} )^{2}
+i\hat{\rho}\left[\dot{\rho}-\sum_{i}\nabla_{i}\left(
D(\rho )\nabla_{i}\frac{\delta{\cal H}_{\rho}}{\delta \rho}\right)\right]\right]
\nonumber
\ee
where $\hat{\rho}$ is the usual field conjugate to $\rho$.
The theory from this point of view will be discussed elsewhere.

We will analyze this problem using the Fokker-Planck (FP) 
formalism\cite{NESM}.
In this approach the density time correlation function
is given by
\be
C({\bf x}_{1}-{\bf x}_{2},t) =
\langle \delta\rho ({\bf x}_{2})e^{-\tilde{D}_{\rho}t}
\delta\rho ({\bf x}_{1})\rangle
\nonumber
\ee
where the average is an equilibrium average over
the static distribution $e^{-\beta {\cal H}_{\rho}}/Z$ and
the adjoint FP operator is defined by
\be
\tilde{D}_{\rho}=\int d^{d}x \int d^{d}y
\left[\frac{\delta {\cal H}_{\rho}}{\delta \rho ({\bf x})}
-k_{B}T\frac{\delta}{\delta \rho ({\bf x})}\right]
\Gamma_{\rho}({\bf x},{\bf y})
\frac{\delta }{\delta \rho ({\bf y})}
~~~.
\nonumber
\ee
where
\be
\Gamma_{\rho}({\bf x},{\bf y})=
\nabla_{x}\cdot \nabla_{y}
\left(D(\rho ({\bf x}) )
\delta \left({\bf x}-{\bf y}\right)
\right)
~~~.
\nonumber
\ee
It is technically important\cite{tech} that
\be
\int d^{d}z \frac{\delta }{\delta \rho ({\bf z})}
\Gamma_{\rho}({\bf z},{\bf x})=0
\nonumber
~~~.
\ee

We study a model with a simple physical choice for $D(\rho )$:
\be
D(\rho )=D_{0}\theta (\rho_{c}-\rho )\frac{(\rho_{c}-\rho )}{\rho_{c}}
\label{eq:1}
\ee
where $D_{0}$,  introduced in Eq.(\ref{eq:2}),
and $\rho_{c}$ are positive parameters.  This form reflects the
geometrical fact that as the density increases it is more difficult for
particles to move and there is some local density $\rho_{c}$ above which
a particle is stuck.  This is physically similar to spin-facilitated
models where mobility is diminished if the local environment is blocked.
We call this model the hindered diffusion model.
More realistic models involve putting in more static structure
and coupling the density to the other slow variables\cite{rick} in the system.

In order to treat the high density constraint adequately we have organized
the theory using the memory function formalism developed in Refs.
\onlinecite{FRKT}
and \onlinecite{MRT}.
The Laplace-Fourier transform, 
\be
C(k,z)=-i\int_{0}^{\infty}e^{izt}\int d^{d}x_{1}
e^{i{\bf k}\cdot ({\bf x}_{1}-{\bf x}_{2})}
C({\bf x}_{1}-{\bf x}_{2},t)
~~~,
\ee
satisfies the
kinetic equation
\be
[z+K(k,z)]C(k,z)=\tilde{C}(k)
\label{eq:56}
\ee
where
$\tilde{C}(k)=k_{B}T \chi (k)$
is the static density correlation function
and $K(k,z)=K^{(s)}(k)+K^{(d)}(k,z)$ is the memory function.
We consider first
the approximation
where we keep only the static part of the memory function: $K^{(s)}$.
Formally $K^{(s)}$
is of order $D_{0}$ while the dynamic part $K^{(d)}(k,z)$
is of order $D_{0}^{2}$.
We will discuss approximations for $K^{(d)}$, including mode coupling contributions,
later.

The static part of the memory function is given, without approximation, by
\be
K^{(s)}(k)=i\beta^{-1}k^{2}\bar{D}\tilde{C}^{-1}(k)
=ik^{2}\bar{D}\chi^{-1} (k)\equiv i\Gamma_{0}(k)
~~~.
\label{eq:65}
\ee
where $\bar{D}=\langle D(\rho )\rangle$ is the average diffusion
coefficient.
If $\rho_{0}=\langle \rho\rangle$ is
the average uniform density and
$S=\langle (\delta \rho)^{2}\rangle$,
then, because the static fluctuations are Gaussian, we must have for the
singlet probability distribution
\be
P[\sigma ]=\langle \delta (\sigma -\rho (1)\rangle
=\frac{e^{-\frac{1}{2S}(\rho_{0}-\sigma )^{2}}}{\sqrt{2\pi S}}
~~~.
\label{eq:15}
\ee
The parameter $S$ is related to the static structure factor by
\be
S
=\int \frac{d^{d}k}{(2\pi )^{d}}k_{B}T\chi (k)
\label{eq:16}
\ee
and $S$
is proportional to the temperature.  The average 
diffusion coefficient can be evaluated as
\be
\bar{D}
=\frac{\rho_{0}D_{0}}{\rho_{c}\epsilon_{0}2\sqrt{\pi}}
\left[\sqrt{\pi}\epsilon\left( erf (\epsilon ) +erf (\epsilon_{0})\right)
+e^{-\epsilon^{2}}-e^{-\epsilon^{2}_{0}}\right]
\ee
where we have introduced the dimensionless parameters
$\epsilon =\frac{(\rho_{c}-\rho_{0})}{\sqrt{2S}}\approx1/\sqrt{T}$
and
$\epsilon_{0} =\frac{\rho_{0}}{\sqrt{2S}}$.
In the high density $\epsilon < 0$, low temperature limit $|\epsilon |$
and $\epsilon_{0}$ are both large
and we find that
the physical diffusion coefficient is activated in temperature:
\be
\bar{D}=D_{0}\frac{\rho_{0} }{\rho_{c}}\frac{1}{2\sqrt{\pi}\epsilon_{0}}
e^{-\epsilon^{2}}\left(\frac{1}{2\epsilon^{2}}-\frac{3}{4\epsilon^{4}}+\ldots\right)
~~~.
\label{eq:4}
\ee
The activated dynamics results from the fact that if $\rho_{0} >\rho_{c}$,
then one can have motion only if there is a local fluctuation where
$\rho < \rho_{c}$.  At low temperatures such fluctuations will be rare.

Putting Eq.(\ref{eq:65}) back into Eq.(\ref{eq:56}), assuming
$K=K^{(s)}$, and inverting the Laplace
transform gives 
\be
C(k,t)=e^{-k^{2}\bar{D}\chi^{-1} (k)t}\tilde{C}(k)=e^{-\Gamma_{0}(k)t}\tilde{C}(k)
~~~.
\label{eq:66}
\ee
One has dramatic slowing down for $\epsilon < 0$ and large.

The dynamic part of the memory function is the sum of
two pieces:
\be
K^{(d)}(q,z)\tilde{C}(q)=
\Gamma^{(d)}(q;z)=
\bar{\Gamma}(q;z)+
\Gamma_{sub}(q;z)
\nonumber
.
\ee
In coordinate space, the first piece is given by
\be
\bar{\Gamma}({\bf x}_{1}-{\bf x}_{2};z)
=-\langle v({\bf x}_{2})R(z)v({\bf x}_{1})\rangle
\label{eq:24}
\ee
where the {\it current} is defined by
$v({\bf x}_{1})=i\tilde{D}_{\rho}\rho ({\bf x}_{1})$ and 
$R(z)=[z+i\tilde{D}_{\rho}]^{-1}$
is the resolvent operator.
The Fourier transform of the second term,
the {\it subtraction} part, is given by
\be
\Gamma_{sub}({\bf q};z)=
\langle v_{-{\bf q}}R(z)\rho_{{\bf q}}\rangle
C^{-1}({\bf q};z)
\langle \rho_{-{\bf q}}R(z)v_{{\bf q}}\rangle
~~~.
\label{eq:69}
\ee
In carrying out the designated averages it is useful to write the current
in the form
\be
v(1)=-i\sum_{\alpha_{1}}\nabla_{x_{1}}^{\alpha_{1}}\int d\sigma D(\sigma )
g_{\sigma}(1)\tilde{\rho}(1)
~~~.
\label{eq:76}
\ee
where
$g_{\sigma}(1)=\delta (\sigma -\rho (1))$
and we have introduced the field
\be
\tilde{\rho}(1) =\tilde{\rho}_{\alpha_{1}}({\bf x}_{1},t_{1})=
\nabla_{x_{1}}^{\alpha_{1}}\int~d^{d}x_{2}\chi^{-1}
({\bf x}_{1}-{\bf x}_{2})\rho ({\bf x}_{2},t_{1})
~~~.
\nonumber
\ee

In evaluating $\Gamma^{(d)}$
we make the simplest approximation.  We assume that the fields $\rho$
appearing in the averages in Eqs.(\ref{eq:24}) and (\ref{eq:69}), 
can be treated as Gaussian.
Corrections to this approximation
will be discussed elsewhere.  
We need, in the coordinate and time regime, the notation
\be
C(12)=\langle\delta\rho (1)\delta \rho (2)\rangle
\nonumber
\ee
\be
C(1\alpha_{1},2)=\langle\tilde{\rho} (1)\delta \rho (2)\rangle
\nonumber
\ee
\be
C(1\alpha_{1},2\alpha_{2})=\langle\tilde{\rho} (1)\tilde{\rho} (2)\rangle
\nonumber
\ee
After doing the Gaussian averages,
the subtraction term, Eq.(\ref{eq:69}), is given by
\be
\Gamma_{sub}({\bf x}_{1}-{\bf x}_{2};z)
=-\sum_{\alpha_{1}\alpha_{2}}\nabla_{x_{2}}^{\alpha_{2}}\nabla_{x_{1}}^{\alpha_{1}}
C({\bf x}_{1}\alpha_{1},{\bf x}_{2}\alpha_{2};z)\bar{D}^{2}
~~~.
\nonumber
\ee
The more complicated contribution
to the dynamic part of the memory function and is given,
in the time regime, by
$\bar{\Gamma}(12)=-\langle v(2)v(1)\rangle$.
After carrying out the average we find
\be
\bar{\Gamma}(12)=
\sum_{\alpha_{1}\alpha_{2}}\nabla_{x_{1}}^{\alpha_{1}}\nabla_{x_{2}}^{\alpha_{2}}
\Bigg[C(1\alpha_{1},2\alpha_{2})
\nonumber
\ee
\be
+C(1\alpha_{1},2)C(2\alpha_{2},1)
\frac{\partial}{\partial C(12)}\Bigg]W(12)
\label{eq:100}
\ee
where $W(12)$ is the diffusion-diffusion correlation function that
has the power-series representation
\be
W(12)=\langle D(1)D(2)\rangle=
\sum_{\ell =0}^{\infty}\frac{(C(12))^{\ell}}{\ell!}
\bar{D}_{\ell}^{2}
\ee
where
we have introduced the average of the derivatives of the bare
diffusion coefficient:
\be
\bar{D}_{\ell}
=\int d\sigma_{1}P[\sigma_{1}]
\frac{\partial^{\ell}}{\partial \sigma_{1}^{\ell}}D(\sigma_{1})
~~~.
\ee
It is convenient to break $W(12)$ into three pieces
\be
W(12)=\bar{D}^{2}
+\bar{D}_{1}^{2}C(12)+\left(\frac{D_{0}}{\rho_{c}}\right)^{2}
\frac{S}{2\pi}\tilde{W}(f)
~~~,
\label{eq:57}
\ee
where the  remaining sum given by
$\tilde{W}(f)$ can be rewritten as an integral
\be
\tilde{W}(f)=
\int_{0}^{sin^{-1}f}dz(f-sin~z)
e^{-\frac{2\epsilon ^{2}}{(1+sin~z)}}
~~~.
\label{eq:188}
\ee
where $f(12)=C(12)/S$.

We can associate distinct terms in $\Gamma^{(d)}$ with the three terms 
in $W(12)$ given in Eq.(\ref{eq:57}).  The $\bar{D}^{2}$ term
can be shown to be equal to $-\bar{\Gamma}_{sub}(12)$ and cancels
the subtraction.  The $\bar{D}_{1}^{2}$ term gives
\be
\Gamma_{MC}^{(d)}(q,t)=\frac{\bar{D}_{1}^{2}}{2}
\int \frac{d^{d}k}{(2\pi )^{d}}
C({\bf k},t)C({\bf q}-{\bf k},t)
\nonumber
\ee
\be
\left[{\bf q}\cdot[{\bf k}\chi^{-1} (k)
+({\bf q}-{\bf k})\chi^{-1} ({\bf q}-{\bf k})]\right]^{2}
\label{eq:112}
\ee
and we have the standard mode-coupling contribution.
The term proportional to $\tilde{W}$ in $W(12)$ gives rise to the
contribution
\be
\Gamma_{BMC}^{(d)}({\bf q},t)
=q^{2}\int\frac{d^{d}k}{(2\pi )^{d}}
\left(\frac{D_{0}}{\rho_{c}}\right)^{2}\frac{S^{2}}{2\pi}
\nonumber
\ee
\be
\times
\chi^{-2}(k){\bf k}\cdot{\bf q}f(k,t)\tilde{W}({\bf q}-{\bf k})
\nonumber
\ee
\be
-\left(\frac{D_{0}}{\rho_{c}}\right)^{2}\frac{S^{2}}{2\pi}
\int\frac{d^{d}k_{1}}{(2\pi )^{d}}
\frac{d^{d}k_{2}}{(2\pi )^{d}}\frac{d^{d}k_{3}}{(2\pi )^{d}}
\tilde{W}'(k_{3})f(k_{1},t)f(k_{2},t)
\nonumber
\ee
\be
\times\delta ({\bf q}-{\bf k}_{1}-{\bf k}_{2}-{\bf k}_{3})
{\bf k}_{1}\cdot{\bf q}~{\bf k}_{2}\cdot{\bf q}
\frac{1}{2}\left(\chi^{-1}(k_{2})-\chi^{-1}(k_{1})\right)^{2}
\nonumber
\ee
where $\tilde{W}(k)$ is the Fourier transform of
$\tilde{W}(f)$ given by Eq.(\ref{eq:188}) and
$\tilde{W}'(k)$ is the Fourier transform of
$\frac{\partial}{\partial f} \tilde{W}(f)$.
The dynamic part
of the memory function is given by $K^{(d)}(q,t)=
\Gamma^{(d)}(q,t)/\tilde{C}(q)$ and the static part by Eq.(\ref{eq:65})
These results go back into the kinetic equation given by Eq.(\ref{eq:56}).
It turns out that we need to make some additional manipulations before
looking for a solution for $C(q,t)$.

If we take the inverse Laplace transform of the kinetic Eq(\ref{eq:56}),
we find
\be
\dot{C}(t)-iK^{(s)}C(t)-\int_{0}^{t}ds~K^{(d)}(t-s)C(s)=0
~~~.
\label{eq:125}
\ee
It is easy to see that Eq.(\ref{eq:125}) with $K^{(d)}(t)$ given by
$\Gamma^{(d)}(t)/\tilde{C}$, does not lead to an ergodic-nonergodic transition,
instead numerical solutions blow up.
Kawasaki\cite{KKA} suggested that the kinetic equation,
Eq.(\ref{eq:56}),
be rewritten in the form
\be
\left(z+\frac{K^{(s)}}{1+K^{(s)}N(z)}\right)C(z)=\tilde{C}
\label{eq:123}
~~~.
\ee
Comparing with Eq.(\ref{eq:56}) we can solve for $N(z)$  to obtain
\be
N(z)=-\frac{K^{(d)}(z)}{K^{(s)}(K^{(s)}+K^{(d)}(z))}
\ee
In a perturbation theory calculation where we treat $K^{(d)}(z)$ as small
we have
\be
N(z)=-\frac{K^{(d)}(z)}{(K^{(s)})^{2}}
~~~.
\label{eq:127}
\ee
So $N(z)$ can be determined using perturbation theory.
Next we note that
Eq.(\ref{eq:123}) can be written in the form
\be
(1+K^{(s)}N(z))(zC(z)-\tilde{C})+K^{(s)}C(z)=0
\label{eq:129}
\ee
Taking the inverse Laplace transform gives
\be
\dot{C}(t)=-\Gamma_{0}C(t)-\Gamma_{0}\int_{0}^{t}ds~N(t-s)\dot{C}(s)
\label{eq:128}
\ee
where $\Gamma_{0}=-iK^{(s)}$
sets the time scale.
This equation 
produces
a nonergodic phase for the standard mode coupling kernel.

The last step in determining $N(q,t)$ is to choose 
the static correlation function $\tilde{C}(q)$.
The simplest assumption is that the static correlation function
is a constant up to a cutoff:
$\chi (q)= \frac{1}{r}$
for $q < \Lambda$ and zero for larger wavenumbers.
In this case we have 
\be
\Gamma^{(d)}(q,t)=\frac{\bar{D}_{1}^{2}}{2}(rq^{2})^{2}S^{2}\int_{\Lambda}
\frac{d^{d}k}{(2\pi )^{d}}f(k,t)f({\bf q}-{\bf k},t)
\nonumber
\ee
\be
+\frac{(qr)^{2}}{2\pi}\left(\frac{D_{0}}{\rho_{c}}\right)^{2}
S^{2}
\int_{\Lambda}\frac{d^{d}k}{(2\pi )^{d}}{\bf k}\cdot{\bf q}
f(k,t)\tilde{W}({\bf q}-{\bf k},t)
~~~
\label{eq:52}
\ee
where 
$f(k,t)$ is the Fourier transform of
$f(r,t)=C(r,t)/S$.
$\Gamma^{(d)}(q,t)$ is related to $N(q,t)$ by Eq.(\ref{eq:127}),
which takes the form:
\be
N(q,t)=\frac{\Gamma^{(d)}(q,t)}{q^{4}\bar{D}^{2}r^{2}\tilde{C}}
=g_{0}I_{MC}(q,t)+g_{1}I_{BMC}(q,t)
\label{eq:54}
\ee
where in three dimensions the couplings are given by
\be
g_{0}=\frac{2\pi S^{2}}{\tilde{C}}
\left(\frac{\bar{D}_{1}}{\bar{D}}\right)^{2}
~~~;
g_{1}=\frac{4\pi S^{3}}{\tilde{C}}
\left(\frac{\bar{D}_{2}}{\bar{D}}\right)^{2}
~~~.
\nonumber
\ee
The functions $I_{MC}(q,t)$ and $I_{BMC}(q,t)$ follow from
Eqs.(\ref{eq:52}) and (\ref{eq:54}) and the definitions of
$g_{0}$ and $g_{1}$.  Because of the ratios of $\bar{D}_{\ell}$
the $g$'s are not activated in temperature.

The set of equations to be solved, Eqs.(\ref{eq:128}) and (\ref{eq:54}),
 is governed by the parameters $\rho_{c}$,
$\rho_{0}$, $S$ and the wavenumber  cutoff $\Lambda$.
The static correlation function can be written as
$\tilde{C}(q)= 6\pi^{2}S/\Lambda^{3}$
,
which follows from Eq.(\ref{eq:16}).
Time is measured in units of
$(\bar{D}r)^{-1}$.

In the dense frozen region where $\epsilon < 0$ and
is large in magnitude we have
$g_{0}=\frac{2\Lambda^{3}\epsilon^{2}}{3\pi}$ and
$g_{1}=\frac{8\Lambda^{3}\epsilon^{4}}{3\pi}$.
which suggests that for large $\epsilon$ there will be an
ergodic-nonergodic transition.

We solved the coupled set of equations numerically.  
Let us first 
fix $\rho_{c}=1.0$ and the momentum cutoff $\Lambda =1.0$
and look at solutions 
$C(q,t)$.
If
we begin with $\rho_{0}=0.5$ and $S=0.1$
we find that $C(q,t)$
decays exponentially with a decay rate, because of the conservation law,
proportional to $q^{2}$.
Thus the kinetics are less
sensitive to  the cutoff as time evolves.
If we increase $S$ to $0.2$ we see little change in the dynamics.  
However, it is
easy to see that the couplings $g_{0}$ and $g_{1}$ increase with
$S$.  While the model is only physically applicable\cite{NEG} to
structural glasses for small $S$, we can look at solutions 
for larger $S$ and we find an ergodic-nonergodic transition (ENE) along the line
$S_{H}=A_{H}(\rho_{H}-\rho_{0})^{1/3}$
where $A_{H}=2.25$ and $\rho_{H}=2.78$.  This holds even for $\rho_{0}
>\rho_{c}$.

In the dense regime $ \rho_{0}> \rho_{c}$ $\epsilon < 0$
we find a line
of ENE transitions for small values of $S$ satisfying the relationship
$S=A(\rho_{0}-\rho_{c})^{2}$ with a good fit for $A=0.245$.
Thus there is a regime where we simultaneously
have activated dynamics and a nonergodic phase.
As $\rho_{0}$ increases beyond $\rho_{c}$, there is a value of
$\rho_{0}\approx \rho_{H}$ where the high $S$ nonergodic branch meets the 
low $S$ branch and for higher $\rho_{0}$ all states are nonergodic.
One can look at the kinetics in the nonergodic regime and obtain a good fit 
to
$C(q,t)/\tilde{C}(q)=f(q)+A_{q}/t$.
The ENE phase separation curves are determined by fitting to this form
and choosing for fixed $\rho_{0}$ and $q=0.4$ that value of $S$ 
that gives $f(q=0.4)=0.5$.

Similar behavior is found for other choices of $\rho_{c}$ and
wavenumbers.

Acknowledgment: I thank Professor S. Das for useful comments.
This work was supported by the National Science Foundation
through the Materials Research Science and Engineering Center through
Grant No. DMR-9808595.


\begin{thebibliography}{99}
\bibitem{coll} Technically the models discussed here are for  colloidal systems
that are simpler than simple fluids since it
is assumed that one can integrate out the momentum density and 
work with a system
described by the density with a diffusive dynamics.
\bibitem{tech} This result is necessary if the equilibrium
probability distribution is stationary in time.
\bibitem{NEG}  If $S$ is not small then there is a significant
probability $P[\rho <0]=\frac{1}{2}(1-erf(\epsilon_{0}))$
that the density is negative.
\end{thebibliography}
\end{document}